# A Conceptual Model for Bidirectional Service, Information and Product Quality in an IS Outsourcing Collaboration Environment

Subrata Chakrabarty

*Mays Business School, Texas A&M University*

schakrabarty@mays.tamu.edu

## Abstract

*This paper advances theory on the process of collaboration between entities and its implications on the quality of services, information, and/or products (SIPs) that the collaborating entities provide to each other. It investigates the scenario of outsourced IS projects (such as custom software development) where the extent of collaboration between a client and vendor is high. Using the social exchange theory, the proposed conceptual model tries to establish the "bidirectional" nature of SIP quality in a collaborative environment, where the SIPs exchanged are possibly "dependent" on each other, and if any entity wishes to receive high SIP quality then it should make efforts to provide high SIP quality in return too. Furthermore, it advocates increasing efforts to link financial stakes (tangible or intangible monetary benefits or risks) to the quality of SIP being continuously exchanged throughout the project life-cycle.*

## 1. Introduction

Service quality, information quality, and product quality have most often been assumed to be unidirectional (one-way delivery); i.e., the customer simply purchases services, information and/or products from the provider [3, 4, 7, 11, 12, 14, 15, 20]. However, as highlighted in later sections, in the collaboration environment of IS outsourcing, the collaborating entities (namely clients and vendors) receive various forms of services, information and/or products (SIPs) from each other (that is, bidirectional or two-way exchanges of SIPs). Often, the service, information and/or product (SIP) an entity receives from its collaborator serves as a significant input to the SIP that the entity will return to its collaborator. In other words, the SIP any entity receives from its collaborator may be "dependent" on the service that it had earlier provided to the collaborator. The term "bidirectional" is used to describe the phenomenon where collaborating entities (e.g., a client and a vendor) provide/receive SIPs (services, information and/or products) to/from each other in a collaboration project (such as an outsourced custom software development project).

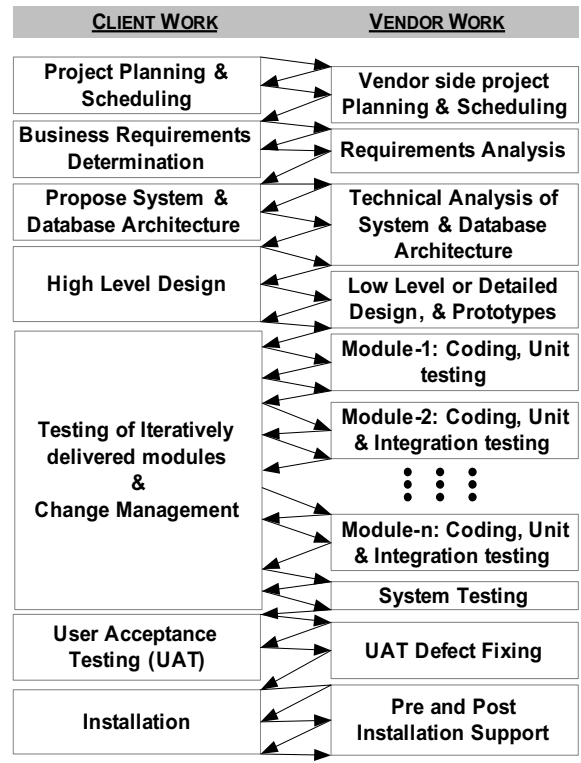

**Figure 1. Example: continuous bidirectional exchanges in custom software development**

For example, let us consider a collaboration environment where a client decides to selectively outsource certain activities of a large custom software development project to a vendor. As shown in figure-1, the client comes up with the initial overall project plan, on the basis of which the vendor comes up with a detailed project plan for the activities to be performed at the vendor site, and also offers suggestions for changes needed in the overall project plan (if any). The client and vendor will of course continuously interact and discuss possible options and alternatives for the project plan, until the plan is finalized. Such continuous interactions and exchanges will be prevalent throughout the remaining phases of the software development life cycle, with each entity (client and vendor) receiving inputs from the other for their respective tasks. After the project planning is completed, the client would primarily provide the business-requirements reports, propose system and database architecture, and suggest a high level design to the vendor, while the vendor would perform requirements analysis, technically analyze of the proposed system and database architecture, and suggest detailed or low-level designs and prototypes. Further, the large software project would be divided into smaller modules, and each module would have to be coded, unit tested, and integration tested by the vendor before delivery to the client. The client would also test each of the modules being iteratively delivered by the vendor to verify that the business requirements and design needs are being met, and return the test result reports to the vendor for suitable defect fixing. The final system would then by tested by the vendor, before the client arranges for user acceptance testing of the software. Additionally, each activity in the software development life cycle described above would involve significant communication, negotiation, detailed documentation, and report generation. Hence, there is a continuous two-way or "bidirectional" exchange between the entities (client and vendor). Also, an activity that has to be performed by one entity may be "dependent" upon the other entity's previously performed activity.

## 1.1. Service, Product and Information Quality

Moving on from the example in the previous section, it is proposed that the collaboration would broadly involve the exchange of the following three attributes:

(1) **Services**: Since tangible cues are much fewer in *services* (and are often limited to the service provider's physical facilities, equipment and personnel), the receiver of the service has to rely on intangible attributes like *reliability, responsiveness, assurance,* and *empathy* associated with the delivered service, and therefore evaluate the "process" of service delivery (or *service quality*) and not solely the outcome of the service (which may even be a product or just information) [19, 20, 21, 22].

(2) **Information**: Significant amounts of "*information*" (such as project status/schedule information, or business and technical requirements information) is exchanged in a collaboration environment, and the associated "*information quality*" would be possibly perceived by the receiver based on attributes such as *reliability, relevance, accuracy, precision, completeness, currency* and *timeliness of the information* [11, p.183; 18, p.44; 24, p.93]. Information is exchanged both formally (documents, reports, emails, letters, meetings, teleconferences …) and informally (discussing work near the water cooler, phone calls, chat, gossip, …) between collaborating personnel. Information may be conveyed through various modes of communication such verbal, visual, or written. The information may be directly associated with the services or products being exchanged, or may be independent information.

(3) **Products**: Tangible intermediate or final "*products*" (such as reports specific to systems/modules, code/programs, prototypes, requirement documents, design documents, test result documents, project status documents, etc…) are delivered by one entity to another, and the "*product quality*" for each of which would be possibly perceived by the receiver based on attributes such as *performance, features, reliability, conformance, durability, serviceability, aesthetics, or overall quality* [10, 20]. The philosophy of quality as "zero defects – doing it right the first time" [20, p.41] or as "conformance to requirements" [5] are targeted towards the quality of goods or products. The quality of products such as software codes or programs can be assessed by reviewing the software code for adherence to coding standards, and by running and testing the code. The quality of the test result documents can be assessed by verifying the comprehensive coverage of the tests, verifying adherence to documentation standards, and finally the effectiveness in finding defects. The quality of design documents can be assessed by validating if the design meets the requirements, if the design documentation follows documentation standards, and finally if the design is easily comprehensible to the software developers. The quality requirements can be assessed by their conceptual clarity and feasibility (within the technology and business constraints), and if the requirements specification documentation standards are followed.

Furthermore, both the provider and receiver of the SIPs affect the process since the receiver's input becomes critical to the quality of SIP delivered by the provider. For example, the barber's service quality would be affected by the customer's description of how the haircut should look (that is, customer's information quality), the doctor's service quality would be affected by the patient's description of symptoms, and the quality of software developed by an IS vendor would be affected by their client's clarity in describing and documenting requirements. [19, 20, 21, 22].

## 1.2. Collaboration Issues in IS outsourcing

The decision to outsource information systems (IS) related work such as custom development of software is growing because organizations need diverse and high-quality IS skills at low costs that enables it to survive and excel in the rapidly changing external environment [1, 3, 4, 7, 16]. Though clients often selectively outsource certain functions of the custom software development project [17], they put the entire onus of the success of the project on the vendor. Clients sometimes fail to realize the importance of "collaboration" in a collaboration project, and wrongly assume that since they are paying the vendor for certain services, information or products, the entire responsibility of the success or failure of the IS project is on the vendor. Hence, vendors are sometimes made scapegoats for a failed project that either slipped in schedule or failed to deliver the client's desired functionality [16, 23]. It is important to realize that for a collaboration project to be successful, the onus for the success of the project should be on the shoulders of each of the collaborators, i.e., both the client and the vendor, and not just the vendor which often plays a subservient role to the client's authority.

While it is important for the client to make sure that it truly collaborates with the vendor by providing the vendor with highest possible quality of SIPs (in addition to the required payments), it is important for the vendor to realize that the client's future in the competitive international business environment may be dependent on the successful execution of the outsourced project. Hence every effort should be made by the vendor to garner maximum support and knowledge from the client such that it can deliver high quality SIPs to the client. IS work is the vendor's "core" capability; however, for the client its "core" capability may not lie in IS but instead in other business sectors [1, 4, 16] such as travel, energy, entertainment, or manufacturing (which make use of IS to survive in the competitive marketplace). Hence the vendor's quality of SIPs in the IS project should definitely be of the highest standards, such that the client can learn, imbibe, and replicate the same.

By theorizing on the SIP quality that collaborating entities (i.e., the client and the vendor) receive from each other, we try to establish the relationship and dependency between the SIP quality that each receives from the other. If such a relationship and the existence of "dependency" is established then it would be appropriate to state that if one wants to receive good SIP quality in collaboration projects, then one must be ready to offer high SIP quality in the collaboration too. If it is established that financial stakes (tangible or intangible monetary gains or losses) in providing or receiving SIP quality have a positive influence on the quality of SIP that is exchanged between collaborators, then it would be pertinent to suggest that collaboration projects should increase efforts to link financial stakes to the *continuous* delivery of high SIP quality throughout the lifecycle of the project.

Though the concept of linking financial stakes to the final outcome i.e., "outsourcing success", has already been established [16, 23], the linking of financial stakes to the *continuous* delivery of high SIP quality during an ongoing collaboration project (and not just the final outcome) is seemingly nonexistent. For example, Sparrow [23] explained *benefit-based relationships* by giving the example of the UK government's employment service (the client) and EDS (the vendor), where both the parties (client and vendor) made an upfront investment in a relationship, and thereafter shared both the benefits and the risks by establishing a payment methodology that links EDS's reward to realized outcomes. Similarly, Millar [1994, as cited in 16, pp. 4-5] wrote about *business benefit contracting*, that involves a contractual agreement defining the vendor's contribution to the client in terms of specific benefits to the business and defining the payment the client will make based upon the vendor's ability to deliver those benefits, thereby matching actual costs with actual benefits and sharing the risks. Though clients and vendors are often vocal about the need for *business benefit contracting*, its actual adoption has been a challenge due to the difficulty associated with *measuring* benefits gained by clients and linking them to an increase (or decrease) in vendor's revenue or profit margins [16].

## 1.3. Research Questions

We know that service quality, information quality and product quality are important in a collaboration environment such as IS outsourcing. But the IS outsourcing literature is primarily concerned with their unidirectional nature or one-way transmission, and not the bidirectional nature of the *continuous exchange* of services, information and products that are prevalent in collaboration environments. We do not know how the "dependency" between the SIPs being exchanged in a collaboration environment affects the SIP quality. The influence of "financial stakes" on SIP quality in a collaboration environment is also not known. These are gaps in the literature that need to be researched. This study intends to address the following questions:

*When entities collaborate by continuously exchanging SIPs:*
- *Is there any relationship between the SIP qualities that each entity receives from the other?*
- *Does the dependency between the SIPs being provided and received (exchanged) affect their quality?*
- *Does each entity's financial stakes in SIPs being exchanged affect their quality?*

This "bidirectional" nature of SIP quality is built on a theoretical framework that is based on the social exchange theory and its adaptation to SIP quality. Future research will be aimed at empirically testing (using field studies) the conceptual model (to be proposed in this paper), which attempts to address the mentioned issues.

## 2. Theoretical Development

Wang et al. [25] argue that custom software development projects involve extensive communication where clients and vendors solicit information, requirements and operational procedures. Their model is based on two main assumptions: (1) the negotiation between the entities involved in software development (for e.g., client and vendor) is primarily non-cooperative with each focusing greater on their own self-interests rather than their collaborator's interests, and (2) there is an imbalance in the *information* available or shared, which intensifies the need for greater communication. Furthermore, it is suggested that greater communication, participation and mutual monitoring between the involved entities helps in reducing uncertainties. In this paper, the prevalence of *continuous* exchange of SIPs between the clients and vendors in a custom software development project will be explained using the social exchange theory, which will also corroborate the need for considering SIPs in collaboration environments as having a "bidirectional dependency".

### 2.1. Outsourcing

Due to various factors, organizations (clients) often need to *outsource* work to external entities (vendors) [1, 4, 7, 16]. Hence, when the service, information, and/or product provider is a non-client entity, such as a vendor/supplier, the process is known as *outsourcing*. When the service, information, and/or product provider to the client is a *client-entity* such as a *subsidiary* or the *internal IS department of the client*, it is known as *insourcing*. *Outsourcing* has been defined in ways that espouse a *selective* approach, for example: *"...selectively turning over to a vendor some or all of the IS functions..."* [1, p. 289], *"...contracting of various information systems' sub-functions..."* [3, p. 131], and *"...turn over part or all of an organization's IS functions to external service provider(s)..."* [4, p. 209].

*Selective sourcing* is the practice of outsourcing select IS applications to vendors, while retaining other IS applications in-house [17, pp. 13-14], based on their respective strengths and capabilities (see figure 2).

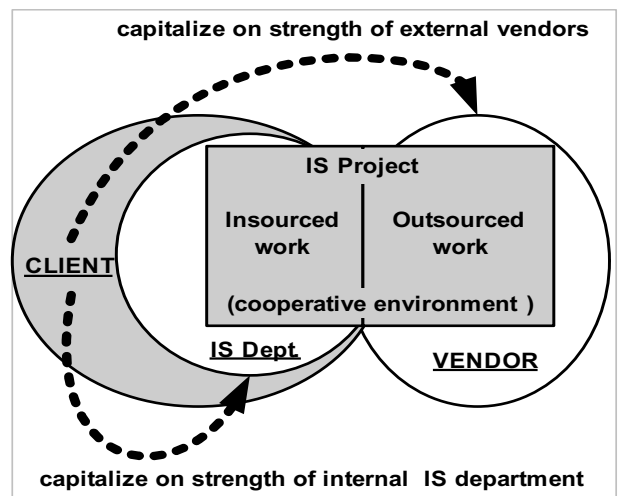

**Figure 2. Selective & Cooperative sourcing**

Outsourcing can be flexible & modular where all the IS functions are broken down into multiple modules, some of which are outsourced and some are retained in-house based on cost analysis, technology and resource needs. In *selective sourcing*, clients outsource between 20 to 60% of the IS budget to vendors (typically around 40%) while still retaining a substantial internal (or in-house) IS department [16, p. 4, pp. 223-224; 5, p. 10]. When a targeted IS activity is performed jointly by the client's internal IS department and the vendor, it is also known as *cooperative sourcing* [Millar, 1994, as cited in 16, pp. 4-5]. *Collaboration* involves a client selectively outsourcing certain functions of a project (such as custom software development) to a vendor, and thereafter both the client and vendor work cooperatively towards the success of the whole project (see figure 2). When clients and vendors collaborate, they exchange services, information and/or products (SIPs).

## 2.2. Social Exchange Theory & Bidirectional SIP Flow

Attempts have been made to apply the social exchange theory to IS outsourcing in the literature to study various relationship, vendor, and contract characteristics [14, 15]. Be it collaboration between individuals, teams, companies or nations, it has always been a "give and take" relationship as established by the social exchange theory. *Social exchange* has been defined by Blau [2] as the *"voluntary actions of individuals that are motivated by the returns they are expected to bring and typically do in fact bring from others"* (see figure 3). Figure-4 illustrates the exchanges in a custom software development project. In addition to paying the vendor, the client has to also provide the vendor with extensive "inputs" (such as requirements, high-level design, business knowledge, and test results of the software coded by the vendor), such that vendor can use these extensive "inputs" to effectively provide the SIPs they are being paid for (such as low-level design of the software, writing of software code, and testing the software being coded).

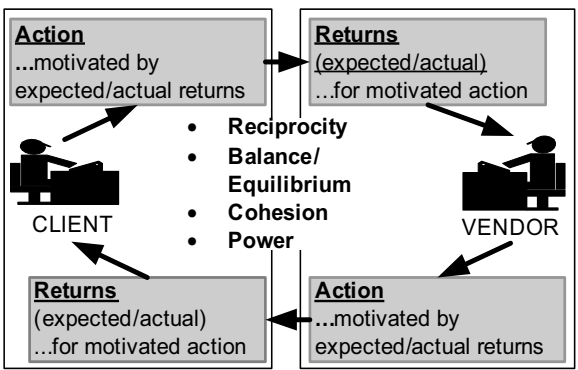

**Figure 3. Social exchange theory**

We characterize the extensive "inputs" that the client gives to a vendor as client-to-vendor "services", "information", or "products" (or appropriate combinations of these) that are used to ensure that the vendor can satisfactorily perform its role. The "client-to-vendor" SIPs and "vendor-to-client" SIPs together comprise the "bidirectional" SIPs in the outsourcing collaboration (see figure 4). Note that the *payments* made by the client to the vendor are distinct from the SIPs, and is later discussed as a component of *"financial stakes"* that effect Bidirectional SIP Quality.

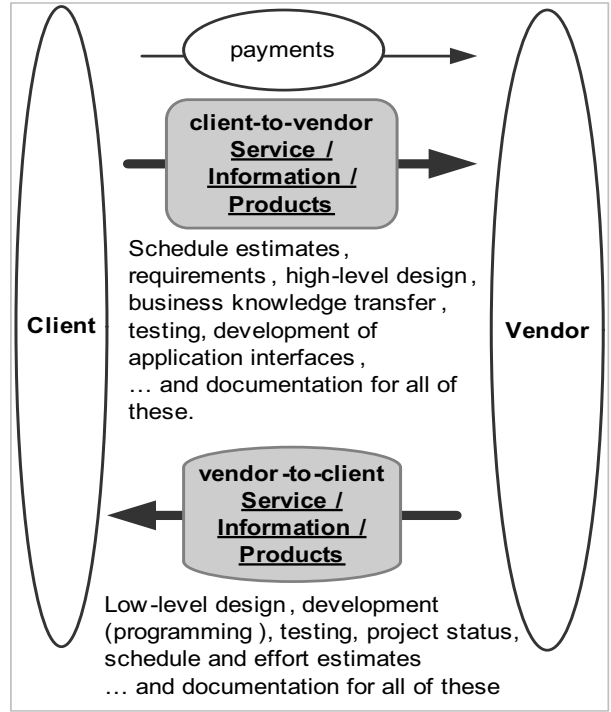

**Figure 4: Bidirectional flow of SIPs**

Emerson [8] noted the following four social-exchange attributes: (1) *Reciprocity* -a mutual exchange as a result of the need to reciprocate the benefits received, (2) *Balance* - an equilibrium or equality in distribution due to mutual dependence between each of the actors in an exchange, (3) *Cohesion* – sticking together when one or more entity runs into a conflict involving the exchange, and (4) *Power* - the amount of monetary influence one can exercise on the other. Section 2.3 will further discuss the relevance of these attributes in forming constructs.

It is important to note that a collaboration environment is not necessarily limited to two entities alone, examples of which are: (1) *multiple-vendor* sourcing (one client, many vendors), (2) *Co-sourcing* (alliance of multiple contract with a single vendor), and (3) *Complex sourcing* (multiple clients and multiple vendors in the same outsourcing contract or alliance) [3; 6, pp. 122-123; 7, pp. 12-13; 9, pp. 1-6]. Though such collaboration arrangements seem to be very complex to handle with respect to the proposed research question, we should recognize that the collaboration in terms of exchange of SIPs between entities is often on a one-to-one basis. That is, it can be assumed that each entity establishes a direct relationship, communicates and deals individually with each of its collaborators. Hence, the to-and-fro flow of SIPs between any two collaborators may be considered to "bidirectional" irrespective of the total number of collaborators.

## 2.3. The Constructs

As discussed earlier, the social exchange theory speaks about actions that are motivated by the returns they are anticipated to bring. On similar lines, the quality of SIPs provided (i.e., the actions) of entities in a collaboration are motivated by the returns they are anticipated to bring. We introduce two constructs that are directly related to the exchange of SIPs:

**(a) *Bidirectional SIP Quality Dependency:*** This construct determines the necessity for an entity to provide its collaborator with high SIP quality, if the entity wants to receive high quality SIPs in return from its collaborator. In other words, if any entity wishes to receive high SIP quality then it should make efforts to provide high SIP quality too. In a collaboration environment, the SIP quality an entity receives from its collaborator may be "*dependent*" on the SIP quality that it provides to the collaborator, and this is termed the entity's "*bidirectional SIP quality dependency*". This is in line with Emerson's [8] social exchange attributes of "reciprocity" (a mutual exchange as a result of the need to reciprocate the benefits received) and "balance" (an equilibrium or equality in distribution due to *mutual dependence* between each of the actors in an exchange).

**(b) *Financial stakes in Bidirectional SIP Quality:*** The payments made (by the client to the vendor) and possible monetary gains and losses are certainly the most obvious and visible "financial stakes". In addition to immediate financial concerns, financial implications of intangibles such as goodwill, trust and future business prospects should also be considered. This is in line with Emerson's [8] social exchange attribute of "*power*" (the amount of monetary influence one can exercise on the other).

The "*financial stakes in bidirectional SIP quality*" construct has two dimensions. The first dimension "*financial stakes in SIP quality provided to collaborator*" determines if there are any financial benefits or costs (financial stakes) in delivering high or low SIP quality to a collaborator, respectively; high benefits for good SIP quality or high costs for poor SIP quality could motivate an entity to deliver high SIP quality such that it increases benefits and reduces costs. The second dimension "*financial stakes in SIP quality received from collaborator*" determines if there is a financial gain or loss if an entity receives good or poor SIP quality respectively; financial gains due to good quality and financial losses due to poor quality of received SIPs will motivate the entity to press the need for high quality from its collaborator that is providing the SIPs.

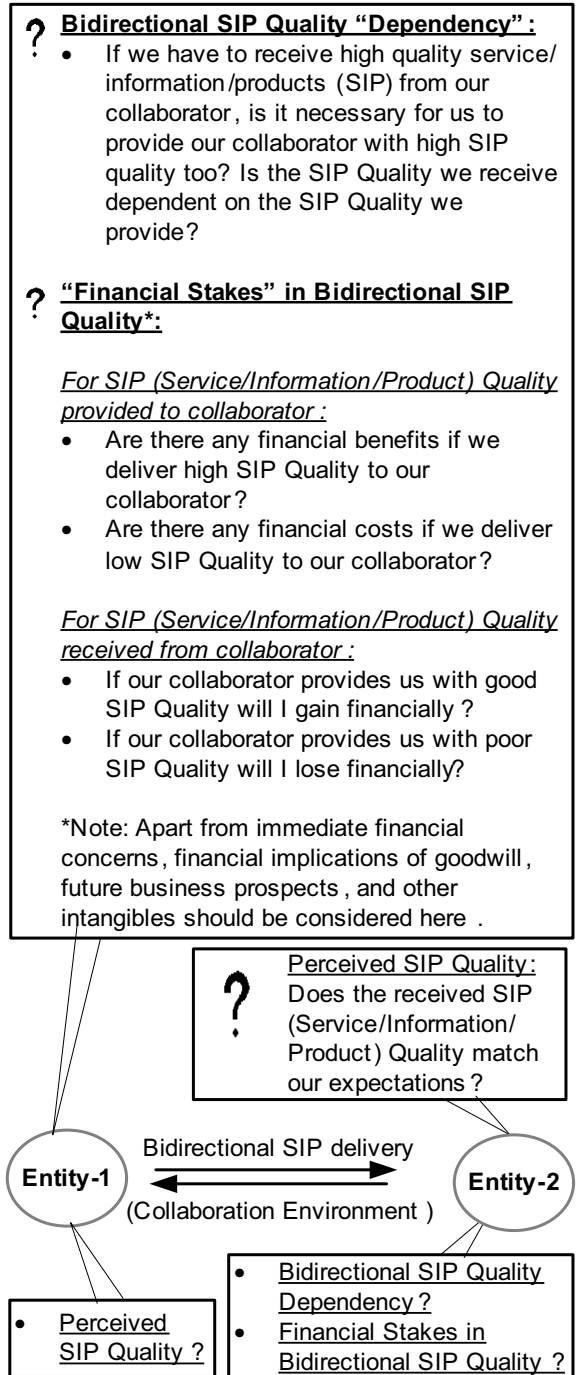

**? Bidirectional SIP Quality "Dependency":**
- If we have to receive high quality service/ information/products (SIP) from our collaborator, is it necessary for us to provide our collaborator with high SIP quality too? Is the SIP Quality we receive dependent on the SIP Quality we provide?

**? "Financial Stakes" in Bidirectional SIP Quality*:**

*For SIP (Service/Information/Product) Quality provided to collaborator:*
- Are there any financial benefits if we deliver high SIP Quality to our collaborator?
- Are there any financial costs if we deliver low SIP Quality to our collaborator?

*For SIP (Service/Information/Product) Quality received from collaborator:*
- If our collaborator provides us with good SIP Quality will I gain financially?
- If our collaborator provides us with poor SIP Quality will I lose financially?

*Note: Apart from immediate financial concerns, financial implications of goodwill, future business prospects, and other intangibles should be considered here.

**?** Perceived SIP Quality: Does the received SIP (Service/Information/ Product) Quality match our expectations?

Bidirectional SIP delivery

Entity-1 ⟷ Entity-2

(Collaboration Environment)

- Perceived SIP Quality?

- Bidirectional SIP Quality Dependency?
- Financial Stakes in Bidirectional SIP Quality?

**Figure 5: Service/Information/Product Quality in collaboration: The questions asked**

Additionally, the construct **"*Perceived SIP Quality*"** which determines if the received SIP quality matches expectations [20, 21, 22], leads to two constructs when adapted to the client-vendor collaboration scenario: (1) "*client-to-vendor*" *SIP quality (perceived by vendor)*, and (2) "*vendor-to-client*" *SIP quality (perceived by client)*, which reflect the bidirectional nature of services within a collaboration framework.

Figure 5 represents each of these constructs in a generic question like format where the collaborating entities self assess their perceptions about the constructs. These questions can form the basis of how the respective constructs can be possibly measured. From figure 5, it is evident that the generic questions asked by both the collaborators are the same. However, when the collaborators are actually given identities (say a particular client or vendor) these questions may then be suitably adapted to address their specific concerns.

## 2.4. Causal effects in Bidirectional SIP Flow

The questions shown in figure-5 when applied to the specific collaboration environment of outsourced IS projects, can lead to a conceptual model and more specific propositions for client-vendor collaborations.

In figures 6, 7, 8, & 9 that follow, the left side contains client related activities & constructs, and the right side contains vendor related activities and constructs. The shaded rounded rectangles denote the actual activities performed, and the associated bold arrows indicate the flow or the order in which the activities are performed. The solid rectangles denote the perceived SIP quality constructs, while the dashed rectangles denote the constructs that effect perceived quality (namely the *dependency* and *financial stakes* constructs). The thin arrows with plus signs indicate a positive influence among constructs.

Figure 6 shows the proposed causal influences when a client provides SIPs to a vendor, and receives SIPs in return from the vendor. In the first and second steps, the client prepares and delivers "client-to-vendor" SIP to the vendor. The vendor's attitude about its perceived quality of the *SIP* that it received from client, affects the way it prepares the SIP (in the third step) that will be delivered (returned) to the client. The vendor's *bidirectional SIP quality dependency* and the vendor's *financial stakes in bidirectional SIP quality* also influences the quality of the "vendor-to-client" SIP delivered in step four.

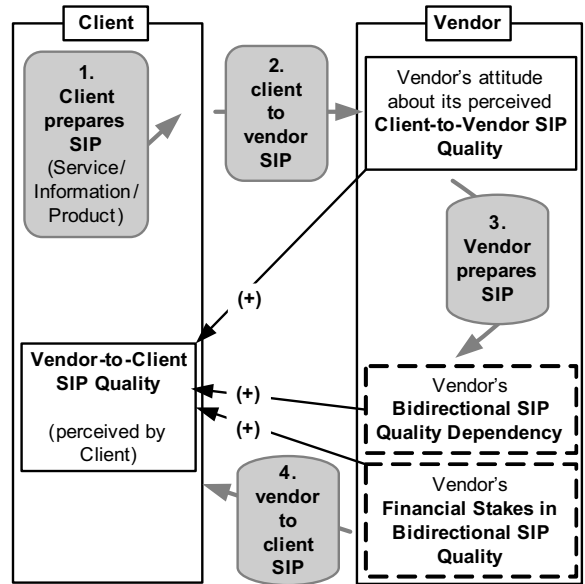

**Figure 6. Causal effects in Bidirectional SIP: Client provides SIP to Vendor and receives SIP in return from Vendor**

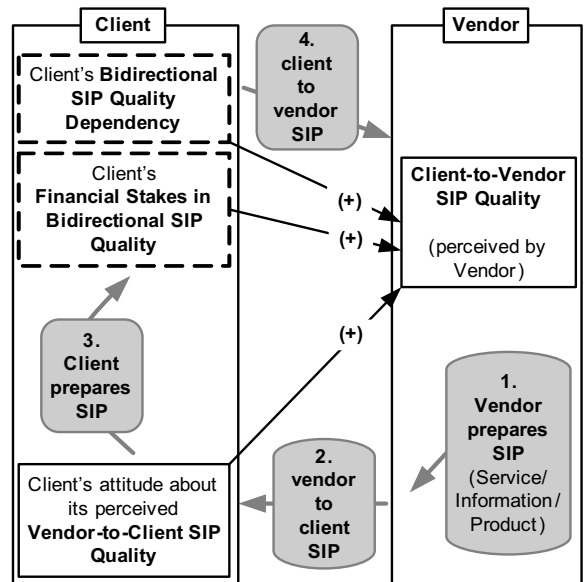

**Figure 7. Causal effects in Bidirectional SIP: Vendor provides SIP to Client and receives SIP in return from Client**

Similarly, figure 7 describes the proposed causal influences when a vendor provides SIPs to a client and receives SIPs in return from the client. In the first and second steps, the vendor prepares and delivers "vendor-to-client" SIP to the client. The client's attitude about its perceived quality of the *SIP* that it received from vendor, affects the way it prepares the SIP (in the third step) that will be delivered (returned) to the vendor. The client's *bidirectional SIP quality dependency* and the client's *financial stakes in bidirectional SIP quality* also influences the quality of the "client-to-vendor" SIP delivered in step four.

Based on the theoretical development thus far, the following are the propositions for possible causal effects:

*For any two entities (client and vendor) that collaborate by providing SIPs to each other ...*

- *...the higher each entity's Bidirectional SIP Quality Dependency, the higher is the SIP quality received by its respective collaborating entity.*
- *...the higher each entity's Financial Stakes in Bidirectional SIP Quality, the higher is the SIP quality received by its respective collaborating entity.*
- *...an entity's attitude about its perceived quality of the SIP received from its collaborating entity will positively influence the quality of SIP that it delivers back to the collaborating entity.*

## 2.5. Correlation and Moderating effects in Bidirectional SIP Flow

A client's attitude about its perceived quality of the SIP received from its vendor can positively influence the quality of SIP that it delivers back to the vendor; and on similar logic, a vendor's attitude about its perceived quality of the SIP received from its client, can positively influence the quality of SIP that it delivers back to the client. Combining these two statements, we propose that there is a positive relationship (correlation) between the "*client-to-vendor*" *SIP-quality (perceived by vendor)* and "*vendor-to-client*" *SIP-quality (perceived by client)*, which is shown in figure 8.

It is further proposed (see figure 8) that each entity's (the client's and the vendor's) perception of *bidirectional SIP quality dependency* and each entity's *financial stakes in bidirectional SIP quality* positively moderates the correlation between the "*client-to-vendor*" *SIP-quality (perceived by vendor)* and "*vendor-to-client*" *SIP-quality (perceived by client)*. This leads to the following propositions on the possible correlation and moderation effects:

*For any two entities (client and vendor) that collaborate by providing SIPs to each other ...*

- *...there is a positive relationship (correlation) between the SIP-quality that each entity receives from the other.*
- *...the higher each entity's Bidirectional SIP Quality Dependency, the stronger is the correlation*

*between the SIP quality that each receives from the other.*

- *..., the higher each entity's Financial Stakes in Bidirectional SIP Quality, the stronger is the correlation between the SIP quality that each receives from the other.*

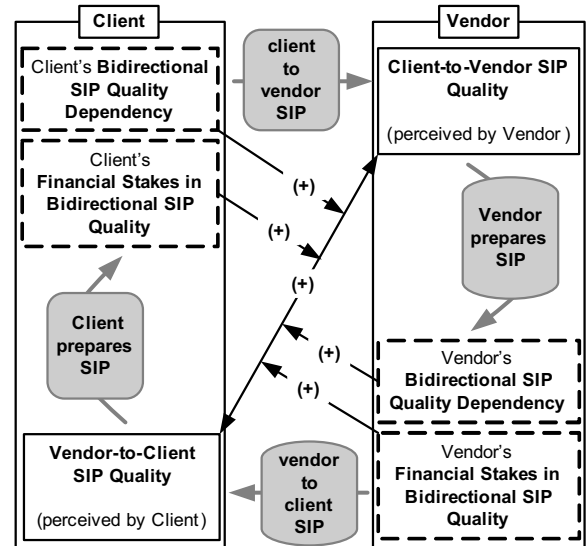

**Figure 8. Correlation in Bidirectional SIPs**

## 3. The Conceptual Model – Bidirectional SIP Qualities

The conceptual model which is the culmination of the theoretical development thus far, shows the possible correlation, causal influences and moderating influences (see figure 9). When "SIP" quality is decomposed into service quality, information quality and product quality, the corresponding constructs would be (1a) "*client-to-vendor*" *service quality (perceived by vendor)*, (1b) "*vendor-to-client*" *service quality (perceived by client)*, (2a) "*client-to-vendor*" *information quality (perceived by vendor)*, (2b) "*vendor-to-client*" *information quality (perceived by client)*, (3a) "*client-to-vendor*" *product quality (perceived by vendor)*, and (3b) "*vendor-to-client*" *product quality (perceived by client)*. The constructs related to "bidirectional SIP quality dependency" and "financial stakes in bidirectional SIP quality" have both causal and moderating influences.

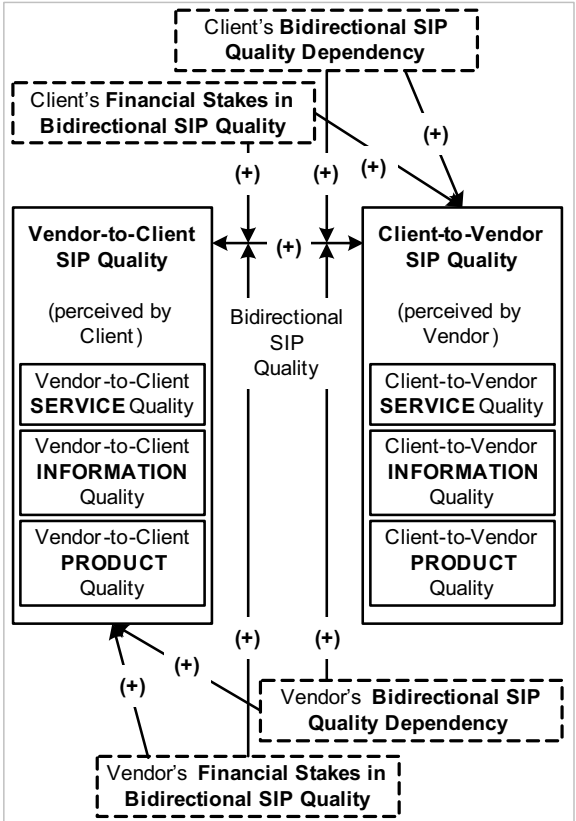

**Figure 9. Conceptual Model: Bidirectional SIP Quality in Client-Vendor collaboration**

The possible *settings* for this conceptual model are outsourced custom software development projects being executed at client and vendor sites. Vendor companies for custom software development like EDS and Accenture are expected to have a large client base, and have a large number of outsourced custom software development projects. The possible *sample* for a field study that can validate this conceptual model can have project managers from both vendor and client sides from each of these collaboration projects.

A plethora of variables and variance is associated with IS outsourcing [1, 4, 7, 18, 25]. The possible variations may arise from project size, age of relationship, relationship/partnership quality, strategic importance, power politics in client-vendor relationships, environmental technological uncertainty, business uncertainty (external forces), discrepancy in support/staff quality, difficulty in monitoring performance, time difference in working hours, differences in culture etc... the list is endless. However, this conceptual model intends to focus purely on the quality aspects in the bidirectional exchange of services, information and/or products between collaborating entities such as clients and vendors. All other factors which are not directly related to the "bidirectional exchange of services, information and/or products" have been kept out to keep this model "simple, neat and clean".

## 3.1. Building on this model – the next step

This model would certainly be valuable when each of the constructs and the relationships are empirically tested in a field study involving IS projects. Additionally, interactions and relationships with other factors not directly related to the "bidirectional exchange of services, information and/or products" (described in the earlier section) may be considered in future research. Also, analysis of bidirectional SIP quality in a collaboration environment would be valuable when SIP quality trends are periodically tracked. A longitudinal study could help establish bidirectional SIP quality as something that may be improved upon. Clients and vendors, for example, would learn about ways to improve bidirectional SIP quality by administering instruments that measure service, information and product quality (for example., the SERVQUAL instrument measures service quality [19, 20, 21, 22]) and an employee survey every month during the duration of a collaboration project, plus systematically soliciting and analyzing suggestions and complaints from its collaborators [22, p.31].

## 4. Conclusion and Implications

The research community which was hitherto concerned with only the unidirectional nature of SIPs, and not the bidirectional nature of SIPs that are prevalent in collaboration environments, will get a new direction for further research after this conceptual model is established. Bidirectional service, information and/or product quality in collaboration environments is unexplored. Knowledge about "why, what, when, or how" *bidirectional dependencies* and *financial stakes* in collaborative environments influence exchange of services, information and/or products will be useful for future research.

By visualizing the SIP quality that collaborating entities (i.e., the client and the vendor) receive from each other as having a "bidirectional dependency", the conceptual model tries to establish a well reasoned relationship between the SIP-quality that each entity receives from the other. If such a relationship is established then it would be appropriate to state that if one wants to receive good SIP quality in collaboration projects, then one must be ready to offer high SIP quality to the collaborators too. If it is established that financial stakes in providing or receiving SIP quality has a positive influence on the quality of SIP that is exchanged between collaborators, then it would be pertinent to suggest that collaboration projects should increase efforts to link financial stakes to service, information and/or product quality.